\documentclass[twocolumn,showpacs,preprintnumbers, 
               superscriptaddress,
               eqsecnum,nofootinbib]{revtex4}
\usepackage{graphicx, fancybox}
\usepackage{amsmath,amssymb}
\usepackage[dvipdfm, colorlinks=true, pdfstartview=FitV, linkcolor=red, citecolor=blue, urlcolor=blue]{hyperref}
\usepackage{color}

\newcommand{\vect}[1]{\mathbf{#1}}

\newcommand{\sgn}{\mathrm{sgn}}
\newcommand{\Slash}[1]{\ooalign{\hfil/\hfil\crcr$#1$}}

\newcommand{\vp}{\vect{p}}
\newcommand{\vk}{\vect{k}}

\newcommand{\vl}{\vect{l}}
\newcommand{\vgamma}{{\boldsymbol \gamma}}

\newcommand{\vB}{\vect{B}}
\newcommand{\vE}{\vect{E}}
\newcommand{\vX}{\vect{X}}
\newcommand{\vj}{\vect{j}}
\newcommand{\vv}{\vect{v}}

\newcommand{\cp}{e}

\newcommand{\Tr}{\mathrm{Tr}}

\newcommand{\comment}[1]{}

\begin{document}

\preprint{RIKEN-QHP-154}

\title{Nonlinear electromagnetic response in quark-gluon plasma}

\author{Daisuke Satow}
\email{daisuke.sato@riken.jp}
\affiliation{Theoretical Research Division, Nishina Center,
RIKEN, Wako 351-0198, Japan}
\affiliation{Department of Physics, Brookhaven National Laboratory, Upton, NY-11973, USA}

\begin{abstract} 
We perform the first systematic study of the nonlinear electromagnetic currents induced by external electromagnetic field in quark-gluon plasma, both in the two cases that the inhomogeneity of electromagnetic field is small (large) so that the collision effect is important (negligible).
In the former case, we list and classify possible components of the currents in a systematic way, and make an order estimate of each component by using the Boltzmann equation in the relaxation time approximation.
In the latter case, we explicitly calculate the quadratic current by using the Vlasov equation, and find that the current generated by the chiral magnetic effect and the quadratic current can have the same order of magnitude by using the Kadanoff-Baym equation.
We also demonstrate this property by using a possible configuration of electromagnetic field realized in heavy ion collision.
\end{abstract} 

\date{\today}


\maketitle

\section{Introduction}
\label{sec:intro}

When non-central collision occurs in heavy ion collision (HIC) experiment, it is expected that strong electric ($\vE$) and magnetic fields ($\vB$) are generated~\cite{elemag-HIC}.
Such fields would induce the electromagnetic and axial current, and these currents contain information on the properties of the medium, which is quark-gluon plasma at temperature $T$.
The simplest components of these currents are, Ohmic current and the current generated by the chiral magnetic effect (CME)~\cite{Fukushima:2008xe, KW, Satow:2014lva}, which are linear in terms of the electric/magnetic field, and local.
The effect of these currents has been broadly discussed theoretically~\cite{elemag-effectOnHIC, elemag-HIC,KW, Hirono:2012rt, Hongo:2013cqa, Gursoy:2014aka} and experimentally~\cite{Abelev:2009ac}.
However, when the electromagnetic field becomes strong enough, it is likely that higher order components of the current in terms of the field~\cite{Buividovich:2010tn, Gursoy:2014aka} are not negligible compared with the linear component.
Also, the assumption of locality\footnote{Here locality of current means that the current at point $X$ does not depend on electromagnetic field at other point (For example, see Eq.~(\ref{eq:Boltzmann-result-1st})).
By contrast, nonlocal current depends on electromagnetic field at other point, such as the current in Eq.~(\ref{eq:HTL-current}).}
 becomes invalid when the inhomogeneity of the electromagnetic field is so large that the collision effect becomes negligible~\cite{Blaizot:2001nr, Blaizot:1992gn}.
In fact, as we will see in Sec.~\ref{ssc:HIC-vlasov}, it can be possible that the both possibilities are realized in HIC. 
Nevertheless, the nonlocal and the higher order components of the current have not well investigated systematically.
For this reason, It is an interesting task to analyze what kind of current exists, and which component becomes dominant in HIC, in which the inhomogeneous and strong electromagnetic field is expected to be generated, in a systematic way.

In this paper, we analyze the linear and mainly quadratic components of the current in terms of the external electromagnetic field in the quark-gluon plasma, systematically with HIC in mind.
There are two reasons why we focus on the quadratic component and do not consider other components that are higher order than the quadratic one:
One is that, as we will see later, the quadratic component is the most sensitive term to the chemical potential $\mu$ when $\mu/T$ is not so large, which is realized in HIC.
The other is that, as will be seen in Sec.~\ref{sec:KB}, the components that is higher than quadratic one, e.g., cubic or quartic, do not appear if we truncate a systematic expansion, which is called gradient expansion, at the next-to-leading order (NLO). 

We work in the following two regimes:
One is that the inhomogeneity of the electromagnetic field is so small that the collision effect can not be neglected, which is treated in Sec.~\ref{sec:collision}. 
In this case, the current is local, so we can list all the possible form of the current, some of which are found to be forbidden by discussing the charge conjugation and the parity property.
We also calculate the linear and the quadratic currents explicitly by using the Boltzmann equation in the relaxation time approximation, to make an order estimate of each component of the current.
The other regime, in which the inhomogeneity of the electromagnetic field is large so that the collision effect is negligible and the current is nonlocal, is analyzed in Sec.~\ref{sec:no-collision}.
In that section, we calculate the quadratic current explicitly with the Vlasov equation.
We also systematically calculate the current by applying the gradient expansion to the Kadanoff-Baym equation.
As a result, we show that the quadratic current at the NLO in the gradient expansion agrees with the one calculated with the Vlasov equation, while it has been known that the calculation at the leading order (LO) reproduces~\cite{Blaizot:1992gn, Blaizot:2001nr} the result of the hard thermal/dense loop (HTL/HDL) approximation~\cite{Frenkel:1989br, Altherr:1992mf}, and the linear current at the NLO is equal to the CME current~\cite{Chen:2012ca, Son:2012zy, Son:2012wh}.
We also show that at the NLO order, the currents that are higher than quadratic current do not appear.
We demonstrate that the quadratic current can have the same order of magnitude as that of CME current, by using a possible field configuration realized in HIC in Sec.~\ref{ssc:HIC-vlasov}.
We summarize this paper and give concluding remarks in Sec.~\ref{sec:summary}. 
Appendix~\ref{app:CME} is devoted to derivation of the CME current from the Kadanoff-Baym equation.
We derive the expression of the CME current in coordinate space in Appendix~\ref{app:CME-coordinate}.

\section{Electromagnetic field with small inhomogeneity}
\label{sec:collision}
In this section, we consider the case that the inhomogeneity of the electromagnetic field in space/time is so small that we can not neglect the collision effect, and the current becomes local.
In such case, we can list possible form of the current, and pick up the terms allowed by the charge conjugation (C) and parity (P) symmetry.
We note that, in general, the inhomogeneities in space and time are independent quantities, so their orders of magnitude can be different.
In this paper, we assume that they have the same order of magnitude, for simplicity.
We also obtain the linear and quadratic current in terms of the electromagnetic field, and at the zeroth and the first order in terms of inhomogeneity, by using the Boltzmann equation in the relaxation time approximation.
By using its result, we make an order estimate of each term of the currents.

Throughout this section, we consider the case that the chiral chemical potential ($\mu_5$) is zero, due to the following reason:
The time scale of the chiral instability is of order $(g^4T\ln(1/g))^{-1}$~\cite{Akamatsu:2013pjd}, where $g$ is the coupling constant in quantum chromodynamics.
This time scale is much shorter than the time scale we focus on, as will be shown later.
Thus, there appears the instability leading to rapid growing of electromagnetic field in our analysis if $\mu_5$ is finite, so to avoid treating this problem, we consider $\mu_5=0$ case.  
Also, with HIC in mind, we assume that $\mu$ is not much larger than $T$: $\mu\lesssim T$.

\subsection{Classification by using C and P symmetries}
\label{ssc:CPT}

We list and classify the possible components of the currents.
Since we consider the case that the inhomogeneity of the electromagnetic field is small and we are interested in ratio of the orders of magnitude for the components that have different dependence on strength of electromagnetic field, we classify the components in terms of the time/space derivative and strength of the electromagnetic field. 
The vector quantities which can be used to construct current\footnote{We treat $\vE$ and $\vB$ as external fields, so they are regarded as independent quantities here, although they are not if we treat them as dynamical quantities following the Maxwell equations in medium.} are
\begin{align}
\vE,~\vB,~\dot\vE,~\dot\vB,~\nabla.
\end{align}
We are considering the case that the electromagnetic field varies slowly in space and time, so here we neglected the terms that contain more than two space and time derivatives.

First, we list the possible form of the currents that are linear in terms of electromagnetic field.
The possible terms are proportional to $\vE,~\vB,~\dot\vE,~\dot\vB,~\nabla\times\vE,~\nabla\times\vB$.
The first one is the Ohmic current, and the second one is CME current~\cite{Fukushima:2008xe, KW}. 
Some properties of the currents above can be determined by looking at how these quantities transform under the discrete transformations, which is summarized in TABLE~\ref{tab:CPT}.
Since the P property of the current operator is different from those of $\vB$, $\dot\vB$, and $\nabla\times\vE$, these components can not exist and only 
\begin{align}
\label{eq:list-1st}
\vE,~\dot\vE,\nabla\times\vB
\end{align}
 remain as long as $\mu_5$, which violates the P symmetry, is zero.
Also, the C property of the remaining terms is the same as that of the current operator, so these terms can exist in $\mu=0$ case, in which the C symmetry is not broken.
This property implies that, in $T\gg \mu$ case, which is realized in HIC, these terms approximately do not depend on $\mu$.

Next, we discuss the currents that are quadratic in terms of electromagnetic field.
After neglecting the terms whose P property is even, there are the following possible terms:
\begin{align}
\label{eq:list-2nd}
\begin{split}
&\vE\times\vB,~\dot\vE\times\vB,~\dot\vB\times\vE,~
\nabla(\vE^2),~\nabla(\vB^2),~\vE(\nabla\cdot\vE),\\
&(\vE\cdot\nabla)\vE,~\vB(\nabla\cdot\vB),~(\vB\cdot\nabla)\vB.
\end{split}
\end{align}
The C property of these quantities is different from that of the current operator, so these terms should vanish in $\mu=0$ case, in which the C symmetry exists.
Therefore, in $T\gg \mu$ case, these components are expected to be proportional to $\mu$.
This property suggests that the quadratic currents is the most sensitive to $\mu$, when $T\gg\mu$.

\begin{table}[t]
\caption{CP properties of relevant quantities.
$+1$ ($-1$) means even (odd) under a discrete transformation.}
\label{tab:CPT}
\begin{tabular}{l c c c c c c c }
\hline
  &$\vE$ & $\vB$ & $\dot\vE$ &$\dot\vB$ & $\nabla$ & $\vj$\\ \hline \hline
C& -1&-1 &-1 & -1&+1 &  -1 \\
P &-1 &+1 &-1 &+1 &-1  & -1\\
\hline
\end{tabular} 
\end{table}

\subsection{Boltzmann equation in relaxation time approximation}
\label{ssc:Boltzmann-relaxation-approximation}

A conventional way to calculate the induced current is to use the Boltzmann equation.
We work in the relaxation time approximation, in which the collision term has a very simple form.
In this approximation, we can not expect that the quantitative behavior of the result obtained from the Boltzmann equation is correctly produced, but its order estimate is expected to be correct.
The Boltzmann equation in that approximation reads~\cite{Blaizot:2001nr}
\begin{align} 
\label{eq:Boltzmann-relaxation_time_approximation}
\begin{split} 
&Dn_{\pm}(\vk, X)
-\tau^{-1} n^{\text{(eq)}}_{\pm}(|\vk|) \\
&~~~= \mp\cp\left(\vE+\vv\times\vB\right)(X) \cdot\nabla_{\vk} n_{\pm}(\vk, X),
\end{split}
\end{align}
where $D\equiv v\cdot\partial_X+\tau^{-1} $, $n_{\pm}(\vk, X)$ is the distribution function for the quark (anti-quark), $n^{\text{(eq)}}_\pm(|\vk|)\equiv [\exp\{\beta(|\vk|\mp\mu)\}+1]^{-1}$ is the distribution function  at equilibrium, $X^\mu\equiv (X_0,\vX)$, and $v^\mu\equiv (1,\vv)$ with $\vv\equiv \vk/|\vk|$.
$\tau$ is called relaxation time, and its order of magnitude is determined by the collision effect.
The order estimate\footnote{In some literatures~\cite{Arnold:2000dr}, it is assumed that the electrons in addition to the quarks exist as charge carriers. 
In such case, the quarks thermalize rapidly than electrons because of their strong interaction, so the dominant contribution to the conductivity comes from the electrons, which leads to $\tau^{-1}\sim e^4T\ln(1/e)$.
We do not consider such case in this paper, but our analysis can be extended to this case by replacing the estimate of the relaxation time as $\tau^{-1}\sim g^4T\ln(1/g)\rightarrow e^4T\ln (1/e)$.
} using the perturbation theory gives $\tau^{-1}\sim g^4T \ln 1/g$~\cite{Hosoya:1983xm}.
Since we focus on the case that the inhomogeneity of the electromagnetic field in space/time is small so that the collision effect is negligible, namely $\partial_X\ll \tau^{-1}$, we see that $\partial_X\ll g^4T \ln 1/g$, which was assumed at the beginning of this section, is justified when $g\ll 1$.  
For simplicity, in this paper we consider an ultrarelativistic fermion whose electromagnetic charge is $\cp$ and that does not have color/flavor structure, and call that particle quark.
It will be straightforward to modify the charge to the real one and to introduce the color/flavor structure.
The induced current is written in terms of the distribution function as
\begin{align}
\label{eq:current-kinetic}
\begin{split} 
\vj(X)&= 2\cp\int\frac{d^3\vk}{(2\pi)^3} \vv (n_{+}(\vk, X)-n_{-}(\vk, X)),
\end{split}
\end{align}
where the factor $2$ comes from the spin degeneracy.

To obtain the induced current, we expand Eq.~(\ref{eq:Boltzmann-relaxation_time_approximation}) in terms of $\vE$ and $\vB$:
First, we expand the distribution function as $n=n^{\text{(eq)}}+\delta n^1+\delta n^2+{\cal O}(F^3_{\mu\nu})$, where $\delta n^1$ ($\delta n^2$) is linear (quadratic) in terms of $F^{\mu\nu}$.
By using this form, the first order terms in the Boltzmann equation read
\begin{align} 
\label{eq:1st-Boltzmann}
\begin{split} 
&D\delta n^1_{\pm}(\vk, X) 
= \mp\cp \vE(X) \cdot\vv n'{}^{\text{(eq)}}_{\pm}(|\vk|).
\end{split}
\end{align}
We see that the magnetic field vanishes from the equation due to isotropy of the distribution function at equilibrium.
The current at the first order is 
\begin{align}
\label{eq:Boltzmann-result-1st}
\begin{split}
\vj_1(X)&= 
2\cp\int\frac{d^3\vk}{(2\pi)^3} \vv (\delta n^1_{+}(\vk, X)-\delta n^1_{-}(\vk, X))\\
&\simeq -\frac{\cp^2\tau}{\pi^2}\int\frac{d\varOmega}{4\pi} \vv (1-\tau \partial_T) \vE(X) \cdot\vv \\ 
&~~~\times\int^\infty_0 d|\vk||\vk|^2(  n'{}^{\text{(eq)}}_{+}(|\vk|)+ n'{}^{\text{(eq)}}_{-}(|\vk|)) \\
&= \frac{\tau}{3}m^2_D (\vE(X)-\tau \dot\vE(X))  ,
\end{split}
\end{align}
where $m_D\equiv e\sqrt{T^2/3+\mu^2/{\pi^2}}$ is the Debye mass.
We emphasize that we expanded in terms of $\tau \partial_X$, $D^{-1}\simeq \tau(1-\tau v\cdot\partial_X)$, by using $\partial_X\ll \tau^{-1}$.
The term that is proportional to $\vE$ is the Ohmic current, in which the conductivity $\sigma_e$ is given by 
\begin{align}
\label{eq:conductivity-tau}
\sigma_e=\frac{\tau m^2_D}{3}.
\end{align}  
This term is of order $e^2\tau T^2F_{\mu\nu}$ while the second term in the right-hand side is of order $e^2\tau^2 T^2 \partial_X F_{\mu\nu}$.
We see that all the linear terms allowed by symmetry, Eq.~(\ref{eq:list-1st}), have been obtained, except for the $\nabla\times \vB$ term.
The reason of the absence of such term can be traced back to the isotropy of the distribution function at equilibrium, as can be seen from Eq.~(\ref{eq:1st-Boltzmann}).

At the second order in terms of electromagnetic field, the Boltzmann equation reads
\begin{align} 
\begin{split} 
D\delta n^2_{\pm}(\vk, X)
&= \cp^2\left(\vE+\vv\times\vB\right)(X) \cdot\nabla_{\vk} \\
&~~~\times D^{-1} \vE(X) \cdot\vv n'{}^{\text{(eq)}}_{\pm}(|\vk|)\\
\delta n^2_{\pm}(\vk, X) 
&\simeq \cp^2\tau^2
\Bigl[\left(\vE+\vv\times\vB\right) \cdot\nabla_{\vk}  \\
&~~~-\tau v\cdot\partial_X\left(\vE+\vv\times\vB\right) \cdot\nabla_{\vk}  \\
&~~~-\tau\left(\vE+\vv\times\vB\right) \cdot\nabla_{\vk} 
 v\cdot\partial_X\Bigr] \\
 &~~~\times\vE \cdot\vv n'{}^{\text{(eq)}}_{\pm}(|\vk|).
\end{split}
\end{align}
Here we have expanded in terms of $\tau \partial_X$.
From Eq.~(\ref{eq:current-kinetic}), the current at zeroth order in terms of $\tau\partial_X$ is
\begin{align}
\label{eq:Boltzmann-result-2nd-0th}
\begin{split}
 \vj_{2}&=2\cp^3\tau^2\int\frac{d^3\vk}{(2\pi)^3} \vv 
\left[\left(\vE+\vv\times\vB\right) \cdot\nabla_{\vk}  \right]\vE \cdot\vv \\
&~~~\times(n'{}^{\text{(eq)}}_{+}(|\vk|)-n'{}^{\text{(eq)}}_{-}(|\vk|))\\
&=  \frac{\cp^3\tau^2\mu}{3\pi^2} \vE\times\vB.
\end{split} 
\end{align}
This component has the same form as the Hall current, and of order $e^3\tau^2\mu (F_{\mu\nu})^2$.
The current at the first order in terms of $\tau\partial_X$ is given by
\begin{align}
\label{eq:Boltzmann-result-2nd-1st}
\begin{split} 
\vj_{2}(X)&= -2\cp^3\tau^3\int\frac{d^3\vk}{(2\pi)^3} \vv 
[v\cdot\partial_X (\vE+\vv\times\vB)\cdot\nabla_{\vk} \\
&~~~+(\vE+\vv\times\vB)\cdot\nabla_{\vk} v\cdot\partial_X] \vE\cdot\vv \\
&~~~\times(n'{}^{\text{(eq)}}_{+}(|\vk|)-n'{}^{\text{(eq)}}_{-}(|\vk|))\\
&= \frac{e^3\tau^3\mu}{3\pi^2}
\Bigl[\dot\vB\times\vE+2\vB\times\dot\vE+\frac{1}{2}\nabla \vE^2 \\
&~~~-2\vE (\nabla\cdot\vE) -(\vE\cdot\nabla) \vE\Bigr],
\end{split}
\end{align}
which is of order $e^3\tau^3\mu \partial_X (F_{\mu\nu})^2$. 
All these order estimates are summarized in TABLE~\ref{tab:order-estimate-Boltzmann}.
We note that, again, all the terms allowed by the symmetries are obtained, except for the terms that contain two $\vB$ in Eq.~(\ref{eq:list-2nd}).
The reason why such terms do not exist, is the isotropy of the thermal distribution function.
We see that, from TABLE~\ref{tab:order-estimate-Boltzmann}, the ratio of the quadratic current to the linear one is of order $e\tau\mu F_{\mu\nu}/T^2$.
Thus, the quadratic current will have the same order of magnitude as that of the linear one, when the external electromagnetic field is as strong as $F_{\mu\nu}\sim T^2/(e\mu\tau)$.

\begin{figure}[t]
\begin{center} 
\includegraphics[width=0.3\textwidth]{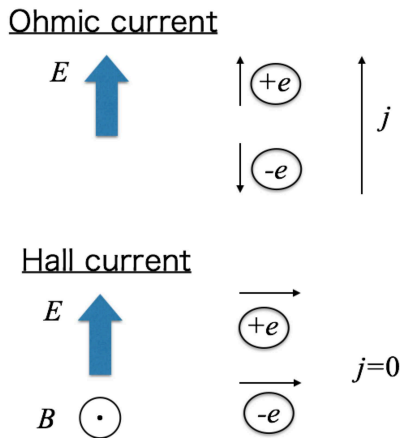}
\caption{Schematic picture of the Ohmic and the Hall currents at $\mu=0$.
The arrows near the quark and the anti-quark show the directions of the forces caused by the electromagnetic field.}
\label{fig:Ohm-Hall}
\end{center}
\end{figure}

We also see that all the second order current, Eqs.~(\ref{eq:Boltzmann-result-2nd-0th}) and (\ref{eq:Boltzmann-result-2nd-1st}), are proportional to $\mu$, while the first order current, Eq.~(\ref{eq:Boltzmann-result-1st}), contain $\mu$-independent terms.
This is consistent with the discussion in the previous subsection.
The physical picture of this property can be explained as follows:
For simplicity, we focus on the Ohmic and the Hall current.
When we consider the linear response of the quark and the anti-quark to the electric field, they move in the opposite direction because they have electric charges with the opposite sign.
Since the current is given by the difference of the quark contribution and the anti-quark one, even in the case that the distribution functions of the two particles at equilibrium are the same, the current exist.
Thus, the Ohmic current is non-zero in the $\mu=0$ case.
By contrast, if the magnetic field acts to these two particles, they feel the Lorenz force with the same signs. 
Therefore, if the distribution functions of the two particles at equilibrium are same, the quark and the anti-quark contributions to the current cancel, so the Hall current does not exist when $\mu=0$.
This explanation is illustrated in Fig.~\ref{fig:Ohm-Hall}.

\begin{table*}[t] 
\caption{Summary of order estimate of the linear and quadratic currents in $\partial_X\ll \tau^{-1}$ case.} 
\label{tab:order-estimate-Boltzmann}
\begin{tabular}{  c c c c }
\hline
 $\vE$ & $\dot\vE$ & $\vB\times\vE$ &$\dot\vB\times\vE, \vB\times\dot\vE, \nabla \vE^2 
, \vE (\nabla\cdot\vE), (\vE\cdot\nabla) \vE$ \\ \hline \hline
$ e^2 \tau T^2F_{\mu\nu}$~~~& $e^2 \tau^2 T^2\partial_XF_{\mu\nu}$~~~ 
& $e^3 \tau^2 \mu (F_{\mu\nu})^2$~~~ & $e^3 \tau^3 \mu \partial_X(F_{\mu\nu})^2$ \\
\hline
\end{tabular} 
\end{table*}

\section{Electromagnetic field with large inhomogeneity}
\label{sec:no-collision}

In this section, we consider the case that the inhomogeneity of the electromagnetic field is large so that we can neglect the collision effect ($\partial_X\gg \tau^{-1}$).
First, we explicitly calculate the current that is quadratic in terms of electromagnetic field explicitly by using the Vlasov equation, which is the kinetic equation without collision term.
Next, we calculate the current induced by external electromagnetic field with the Kadanoff-Baym equation, at NLO of the gradient expansion.
It has been known that, the LO result reproduces~\cite{Blaizot:1992gn} the HTL/HDL current~\cite{Frenkel:1989br, Altherr:1992mf} while the linear current at the NLO agrees with the CME current~\cite{Chen:2012ca, Son:2012zy}.
The quadratic current at NLO is calculated in this paper for the first time, and we show that this current agrees with the one calculated with the Vlasov equation.
Since both of the CME and the quadratic currents are NLO of the gradient expansion, they have the same order of magnitude under the conditions described later.
We also find that the components that are higher than the quadratic one in terms of electromagnetic field, such as cubic and quartic ones, do not exist at the NLO.

In this section, we consider the case $\partial_X\gg \tau^{-1}\sim g^4T \ln(1/g)$, so the time scale we consider is much shorter than that of the chiral instability.
For this reason, we assume that $\mu_5$ is finite in this section.
Also, we calculate the axial current in addition to the vector one for completeness.

\subsection{Vlasov equation}
\label{ssc:Vlasov}

When the inhomogeneity of the electromagnetic field is large enough to neglect the collision effect, the Boltzmann equation is reduced to the Vlasov equation:
If $\partial_X\gg \tau^{-1}$, Eq.~(\ref{eq:Boltzmann-relaxation_time_approximation}) becomes
\begin{align}
\begin{split}
v\cdot \partial_Xn_{\pm L/R}(\vk, X)
&= \mp\cp\left(\vE+\vv\times\vB\right)(X) \cdot\nabla_{\vk} \\
&~~~\times n_{\pm L/R}(\vk, X),
\end{split}
\end{align}
where $n_{+ L/R}$ ($n_{- L/R}$) is the distribution function for the left/right-handed quark (anti-quark).
We separately wrote the equations for the left-handed and the right-handed quark since we have finite $\mu_5$.
The vector and axial currents are given by
\begin{align}
\nonumber
\vj(X)&= \cp\int\frac{d^3\vk}{(2\pi)^3} \vv 
(n_{+L}(\vk, X)-n_{-L}(\vk, X) \\
\label{eq:current-distribution-LR}
&~~~+n_{+R}(\vk, X)-n_{-R}(\vk, X)),\\
\nonumber
\vj^A(X)&= \cp\int\frac{d^3\vk}{(2\pi)^3} \vv
(-n_{+L}(\vk, X)+n_{-L}(\vk, X) \\
\label{eq:current-distribution-LR-axial}
&~~~+n_{+R}(\vk, X)-n_{-R}(\vk, X)).
\end{align}
To obtain the current, we expand the equation in terms of electromagnetic field as $n=n^{\text{(eq)}}+\delta n^1+\delta n^2+{\cal O}(F^3_{\mu\nu})$, where $n^{\text{(eq)}}_{\pm L/R}(|\vk|)\equiv [\exp\{\beta(|\vk|\mp\mu_{L/R})\}+1]^{-1}$.
$\mu_{L/R}= \mu\mp\mu_5$ is the chemical potential for the left/right handed quark.
$\delta n^1$ is determined by the Vlasov equation at the first order, which reads
\begin{align}
\label{eq:Vlasov-linear}
\begin{split}
&v\cdot \partial_X\delta n^1_{\pm L/R}(\vk, X)
= \mp\cp\vE(X) \cdot\nabla_{\vk} n_{\pm L/R}(|\vk|),
\end{split}
\end{align}
whose solution is 
\begin{align}
\begin{split}
\delta n^1_{\pm L/R}(\vk, X)
&= \mp e\int^\infty_0 dt e^{-\eta t}  \vv\cdot\vE(X-vt) \\
&~~~\times n'{}^{{\text{(eq)}}}_{\pm L/R}(|\vk|). 
\end{split} 
\end{align}
Here $\eta$ is an infinitesimal quantity.
It is known~\cite{Blaizot:1992gn, Blaizot:2001nr} that by substituting this expression into Eq.~(\ref{eq:current-distribution-LR}), we reproduce the result of the HTL/HDL approximation~\cite{Frenkel:1989br, Altherr:1992mf}, which read
\begin{align}
\label{eq:HTL-current}
\vj_1(X)&= m^2_D \int\frac{d\varOmega}{4\pi} \vv\int^\infty_0 dt e^{-\eta t}
\vv\cdot\vE(X-vt).
\end{align} 
Here $m^2_D\equiv e^2(T^2/3+(\mu^2+\mu^2_5)/\pi^2)$ is modified from that in $\mu_5=0$ case.
Since the dominant contribution comes from the region $t\sim \partial^{-1}_X$, this current is of order $e^2T^2\partial^{-1}_X F_{\mu\nu}$.
We also see that it is nonlocal.
In the same way, the axial current is shown to be
\begin{align}
\vj^A_1(X)&= \frac{2e^2}{\pi^2}\mu\mu_5 \int\frac{d\varOmega}{4\pi} \vv\int^\infty_0 dt e^{-\eta t}
\vv\cdot\vE(X-vt).
\end{align} 
We note that this current is proportional to $\mu\mu_5$, which is the same parameter dependence as that of the current generated by the chiral electric separation effect~\cite{Huang:2013iia}. 

Also, it is known that the Vlasov equation with the Berry phase term produces~\cite{Chen:2012ca, Son:2012zy, Son:2012wh, Stephanov:2012ki} the following CME current,
\begin{align}
\label{eq:CME-formal}
j^i_{\text{CME}}(X)&= \int d^4Y \varPi^{i\nu}_{R}(X-Y) A_\nu(Y),
\end{align}
where $A_\mu$ is the gauge field, and the retarded polarization tensor for the CME reads~\cite{Son:2012zy}
\begin{align}
\label{eq:CME-Pi}
\begin{split}
\varPi^{ij}_R(p)&= \frac{i\cp^2}{2\pi^2}\mu_5\epsilon^{ijk}\left(1-\frac{p^2_0}{|\vp|^2}\right)p^k \\
&~~~\times\left(1+\frac{p_0}{2|\vp|}\ln \frac{p_0-|\vp|+i\eta}{p_0+|\vp|+i\eta}\right),
\end{split} 
\end{align}
 in momentum space.
We note that this result also can be reproduced by using the Kadanoff-Baym equation.
Since the derivation of Eq.~(\ref{eq:CME-Pi}) with the Kadanoff-Baym equation can not be found in literatures, we write it in Appendix~\ref{app:CME}.
Equation~(\ref{eq:CME-formal}) can be rewritten in terms of electromagnetic field:
\begin{align} 
\label{eq:result-CME}
\begin{split}
\vj_{\text{CME}}(X)&= \frac{e^2}{2\pi^2}\mu_5 \biggl[\vB(X) \\
&~~~+\int\frac{d\varOmega}{4\pi}\int^\infty_0 dt 
e^{-\eta t}\biggl\{\vv\times\dot\vE
-\dot{\vB}\biggr\}(X-vt)\biggr], 
\end{split}
\end{align}
which is of order $e^2\mu_5 F_{\mu\nu}$.
For detail of the derivation of this expression, see Appendix~\ref{app:CME-coordinate}.
The axial current can be obtained by replacing $\mu_5$ with $\mu$ in this expression, which is the current due to the chiral separation effect~\cite{Son:2004tq}. 

Now we focus on the second order response, in which the Vlasov equation becomes
\begin{align}
\label{eq:Vlasov-quadratic}
\begin{split}
v\cdot \partial_X \delta n^2_{\pm L/R}(\vk, X)
&= \mp\cp\left(\vE+\vv\times\vB\right)(X) \cdot\nabla_{\vk}\\
&~~~\times \delta n^1_{\pm L/R}(\vk, X).
\end{split}
\end{align}
By solving this equation, we get
\begin{align}
\begin{split}
&\delta n^2_{\pm L/R}(\vk, X)
=  e^2\int^\infty_0 dt_1\int^\infty_0 dt_2 e^{-\eta(t_1+t_2)}\\
&~~~\times(\vE+\vv\times\vB)(\alpha)\cdot\nabla_{\vk}
  \vv\cdot\vE(\beta)  n'{}^{{\text{(eq)}}}_{\pm L/R}(|\vk|),
\end{split}
\end{align}
where $\alpha\equiv X - v t_1$ and $\beta\equiv X-v(t_1+t_2)$.
The quadratic current is obtained from this expression and Eq.~(\ref{eq:current-distribution-LR}):
\begin{align} 
\label{eq:Vlasov-eqsult-quadratic}
\begin{split}
\vj_2(X)&= \frac{e^3\mu}{\pi^2} \int\frac{d\varOmega}{4\pi}\vv
\int^\infty_0 dt_1\int^\infty_0 dt_2 e^{-\eta(t_1+t_2)}\\
&~~~\times\bigl[E^i(\alpha)\{-E^i(\beta)+3v^iv^jE^j(\beta)\\
&~~~+(t_1+t_2)v^jP^{ik}_T\nabla^k_\vX E^j|_{X=\beta}\} \\
&~~~+(\vv\times\vB(\alpha))^i\bigl\{-E^i(\beta) \\
&~~~+v^j(t_1+t_2)\nabla^i_\vX E^j|_{X=\beta}\bigr\}\bigr],
\end{split}
\end{align}
where $P^{ij}_T\equiv \delta^{ij}-v^i v^j$.
This quantity is of order $e^3\mu\partial^{-2}_X(F_{\mu\nu})^2$.
We note that this expression does not depend on $T$ nor $\mu_5$.
The axial current is obtained by replacing $\mu$ with $\mu_5$ in Eq.~(\ref{eq:Vlasov-eqsult-quadratic}).

\subsection{Kadanoff-Baym equation} 
\label{sec:KB}

The Kadanoff-Baym equation~\cite{Blaizot:1992gn, Blaizot:2001nr, Blaizot:1999xk, Son:2012zy} that is relevant to our study describes the time-evolution of the quark propagator, $S^<(x,y)\equiv \langle \overline{\psi}(y)\psi(x) \rangle $, with $\psi$ ($\overline{\psi}$) is the (anti-)quark field and $\langle ...\rangle$ is the expectation value at nonequilibrium state, which is specified by disturbance characterized by the external photon field ($A^\mu$).
This formalism is a first-principle calculation based on quantum field theory, so even when we use some approximations, what conditions are assumed is clear.
The quark propagator calculated with this formalism is related to the vector and axial current in the following way:
\begin{align}
\label{eq:KB-current-vector}
\vj(x)&= e\Tr[\vgamma S^<(x,x)],\\ 
\label{eq:KB-current-axial}
\vj_A(x)&= e\Tr[\vgamma \gamma_5 S^<(x,x)].
\end{align}
In the presence of external electromagnetic field, the Kadanoff-Baym equation for the quark propagator reads~\cite{Blaizot:2001nr, Blaizot:1992gn} 
\begin{align}
\label{eq:KB}
\begin{split}
&\left(D^2_x-D^{\dagger 2}_y\right)S^<(x,y)
\\
&~~~= -\frac{\cp}{2}\left(F^{\mu\nu}(x) \sigma_{\mu\nu} S^<(x,y)-F^{\mu\nu}(y)S^<(x,y)\sigma_{\mu\nu}\right),
\end{split}
\end{align}
where $D_x\equiv \partial_x+ie A(x)$ is the covariant derivative, $F_{\mu\nu}\equiv \partial_\mu A_\nu-\partial_\nu A_\mu$ is the field strength, and $\sigma_{\mu\nu}\equiv i[\gamma_\mu,\gamma_\nu]/2$.
We neglected the collision effect, which is justified because of $\partial_X\gg \tau^{-1}$~\cite{Blaizot:2001nr, Blaizot:1992gn}.
By introducing $s\equiv x-y$ and $X\equiv (x+y)/2$, the equation becomes
\begin{widetext}
\begin{align}
\label{eq:KB-Xs}
\begin{split} 
&\biggl[2\partial_s\cdot\partial_X 
+ie\biggl\{\left(\left(\partial_s+\frac{\partial_X}{2}\right)\cdot e^{s\cdot\partial_X/2}A(X) 
+\left(-\partial_s+\frac{\partial_X}{2}\right)\cdot e^{-s\cdot\partial_X/2}A(X)\right)\\
&+2e^{s\cdot\partial_X/2}A(X)\cdot\left(\partial_s+\frac{\partial_X}{2}\right) 
+2e^{-s\cdot\partial_X/2} A(X)\cdot \left(-\partial_s+\frac{\partial_X}{2}\right)\biggr\} 
-e^2\{(e^{s\cdot\partial_X/2}A(X))^2
-(e^{-s\cdot\partial_X/2}A(X))^2\}\biggr]S^<(x,y) \\
&= -\frac{e}{2}[(e^{s\cdot\partial_X/2} F^{\mu\nu}(X)-e^{-s\cdot\partial_X/2} F^{\mu\nu}(X))\sigma_{\mu\nu}S^<(x,y)
-(e^{-s\cdot\partial_X/2} F^{\mu\nu}(X))[S^<(x,y),\sigma_{\mu\nu}]].
\end{split}
\end{align}
\end{widetext}
Here we perform the gradient expansion, which is an expansion in terms of $\partial_X/\partial_s$.
Since $\partial_s\sim T$ as will be seen later, we assume $\partial_X\ll T$.
Also, we see that, in the Kadanoff-Baym equation, there is another dimensionless parameter, $eA^\mu/T$.
Since we are not focusing on the region in which the electromagnetic field is so strong that the expansion in terms of electromagnetic field becomes completely useless, we also assume that this quantity is small enough.
Concretely, we assume the following condition:
\begin{align}
\label{eq:KB-condition}
\left(\frac{eA^\mu}{T}\right)^4 \ll \left(\frac{\partial_X}{T}\right)^2 \ll \frac{eA^\mu}{T} \ll 1.
\end{align}
In the derivation of the linearized Vlasov equation~\cite{Blaizot:1992gn}, it was assumed that $eA^\mu\sim \partial_X\sim eT$, so the condition above was satisfied. 
By neglecting the terms that are much smaller than $e^2A^2 \partial_XS^</T$ and $eA \partial^2_XS^</T$, Eq.~(\ref{eq:KB-Xs}) becomes
\begin{align}
\label{eq:KB-gradient-expansion}
\begin{split} 
&2[\partial_s\cdot \partial_X 
+i\cp\left\{A\cdot\partial_X+(\partial_X\cdot A)+(s\cdot \partial_X A_\mu)\partial^\mu_s\right\} \\
&~~~-\cp^2 A^\mu(s\cdot\partial_X A_\mu)]S^<(s,X) \\
&=-\frac{\cp}{2}\biggl((s\cdot\partial_XF^{\mu\nu}) \sigma_{\mu\nu} S^<(s,X) \\
&~~~-\left\{\left(1-\frac{s\cdot\partial_X}{2}\right)F^{\mu\nu}\right\}[S^<(s,X),\sigma_{\mu\nu}]\biggr).
\end{split}
\end{align}
Now we perform the Wigner transformation, which is defined as $f(k,X)\equiv \int d^4 s e^{ik\cdot s}f(s,X)$, where $f$ is an arbitrary function.
After doing this transformation, Eq.~(\ref{eq:KB-gradient-expansion}) reads
\begin{align}
\label{eq:KB-Wigner-transformation}
\begin{split} 
&\left(k-eA\right)^\mu\left[ \partial_{X\mu} +\cp\partial^\nu_k(\partial_{X\nu} A_\mu)
\right]S^<(k,X) \\
&= \frac{\cp}{4}\biggl(-\partial_{k\alpha}(\partial^\alpha_XF^{\mu\nu}) \sigma_{\mu\nu} S^<(k,X)\\
&~~~+i\left\{\left(1-i\frac{\partial_k\cdot\partial_X}{2}\right)F^{\mu\nu}\right\}[S^<(k,X),\sigma_{\mu\nu}]\biggr)
\end{split}
\end{align}
This equation can be rewritten in explicitly gauge invariant form by introducing the gauge covariant Wigner function~\cite{Blaizot:2001nr, Blaizot:1992gn, Blaizot:1999xk}, 
\begin{align}
\begin{split}
\acute{S}^<(s,X)&\equiv U\left(X,X+\frac{s}{2}\right) 
S^<\left(X+\frac{s}{2}, X-\frac{s}{2}\right) \\
&~~~\times U\left(X-\frac{s}{2},X\right),
\end{split}
\end{align}
where $U(x,y)\equiv {\text P}\exp(-ie\int_\gamma dz^\mu A_\mu(z))$ is the Wilson line, with P is the path ordering operator and $\gamma$ is an arbitrary path from $y$ to $x$.
By performing the gradient expansion, the Wilson lines become
\begin{align}
\begin{split}
&U\left(X,X+\frac{s}{2}\right)U\left(X-\frac{s}{2},X\right)
= e^{i\cp s\cdot A(X)} \\
&~~~+{\cal O}\left(\frac{eA \partial_X}{T^3}, \frac{e^2A^2 \partial_X}{T^3}, \frac{e^3A^3}{T^3}\right),
\end{split}
\end{align}
so we have
\begin{align}
S^<(k,X)&= \acute{S}^<(l,X)
\end{align}
up to this order.
Here $l\equiv k-eA$.
By using this relation, Eq.~(\ref{eq:KB-Wigner-transformation}) is written in the following gauge invariant form:
\begin{align}
\label{eq:KB-gauge-covariant}
\begin{split}
&\left[l\cdot \partial_X 
 -\cp l^\mu\partial^\nu_l F_{\mu\nu}\right]\acute{S}^<(l,X) \\
&= \frac{\cp}{4}\biggl(-\partial_{l\alpha}(\partial^\alpha_XF^{\mu\nu}) \sigma_{\mu\nu} \acute{S}^<(l,X) \\
&~~~+i\left\{\left(1+i\frac{\partial_l\cdot\partial_X}{2}\right)F^{\mu\nu}\right\}[\acute{S}^<(l,X),\sigma_{\mu\nu}]\biggr)
\end{split}
\end{align}

Let us obtain $\acute{S}^<$ order by order.
To this end, we expand this quantity as $\acute{S}^<(l,X) =S^{<{\text{(eq)}}}(l)+\delta \acute{S}^{<{\text{LO}}}(l,X)+\delta \acute{S}^{<{\text{NLO}}}(l,X)$, where $\delta \acute{S}^{<{\text{LO}}}$ is of order $S^{<{\text{(eq)}}} eA/T$ and $\delta \acute{S}^{<{\text{NLO}}}$ is of order $S^{<{\text{(eq)}}}\times$max($e^2A^2\partial_X/T^3$, $eA\partial^2_X/T^3$).
The quark propagator at equilibrium is given by
\begin{align}
\label{eq:propagator-eq}
S^{<{\text{(eq)}}}(l)&= \rho^0(l)\left[P_Ln^L(l^0)+P_Rn^R(l^0)\right]\Slash{l} ,
\end{align}
where $\rho^0(l)\equiv  2\pi\sgn(l^0)\delta(l^2)$ is the spectral function of massless particle, $n^{L/R}(l^0)\equiv  [\exp\{\beta(l^0-\mu_{L/R})\}+1]^{-1}$, and $P_{R/L}\equiv(1\pm\gamma_5)/2$.

\subsubsection{leading order} 
The calculation of LO was already performed~\cite{Blaizot:1992gn, Blaizot:2001nr}, but for later convenience, we recapitulate its calculation briefly. 
At the LO, Eq.~(\ref{eq:KB-gauge-covariant}) becomes
\begin{align}
\label{eq:KB-LO}
\begin{split}
&l\cdot \partial_X \delta \acute{S}^{<{\text{LO}}}(l,X)\\
&=\cp F_{\mu\nu} \left(l^\mu\partial^\nu_l S^{< {\text{(eq)}}}(l)
+ \frac{i}{4}[S^{< {\text{(eq)}}}(l),\sigma_{\mu\nu}]\right).
\end{split}
\end{align}
From this equation, we see that $\delta \acute{S}^{<{\text{LO}}}$ has only linear component in terms of $F_{\mu\nu}$. 
By introducing $\delta n^1_{L/R\pm}$ as 
\begin{align}
\label{eq:S-LO-distribution}
\begin{split}
\delta \acute{S}^{<{\text{LO}}}&= 2\pi \delta(l^2)
[\theta(l^0)\{P_L\delta n^1_{L+}+P_R\delta n^1_{R+}\}(\vl,X) \\
&~~~+\theta(-l^0)\{P_L\delta n^1_{L-}+P_R\delta n^1_{R-}\}(-\vl,X)]\Slash{l},
\end{split}
\end{align}
 we see that Eq.~(\ref{eq:KB-LO}) is reduced to the linearized Vlasov equation, Eq.~(\ref{eq:Vlasov-linear}).

To show the equivalence between the linearized Vlasov equation and the Kadanoff-Baym equation at the LO, we also have to show that the expression of the current in terms of the distribution function is the same in the both formalisms.
By using the Wigner-transformed Green's function, Eqs.~(\ref{eq:KB-current-vector}) and (\ref{eq:KB-current-axial}) can be written as
\begin{align}
\nonumber
\vj(x)&= e\int\frac{d^4k}{(2\pi)^4}\Tr[\vgamma S^<(k,x)] \\
\label{eq:current-vector-S}
&= e\int\frac{d^4l}{(2\pi)^4}\Tr[\vgamma \acute{S}^<(l,x)],\\ 
\nonumber
\vj^A(x)&= e\int\frac{d^4k}{(2\pi)^4}\Tr[\vgamma \gamma_5 S^<(k,x)]\\
\label{eq:current-axial-S}
&= e\int\frac{d^4l}{(2\pi)^4}\Tr[\vgamma \gamma_5 \acute{S}^<(l,x)],
\end{align}
where we have used the fact that $\acute{S}$ is obtained from $S$ by shifting the momentum $k$ by $\cp A$.
By substituting Eq.~(\ref{eq:S-LO-distribution}) into Eqs.~(\ref{eq:current-vector-S}) and (\ref{eq:current-axial-S}), we see that these equations are reduced to Eqs.~(\ref{eq:current-distribution-LR}) and (\ref{eq:current-distribution-LR-axial}).
\subsubsection{next-to-leading order} 

Now we calculate the current at the NLO.
At this order, Eq.~(\ref{eq:KB-gauge-covariant}) reads
\begin{align}
\label{eq:KB-NLO}
\begin{split} 
&l\cdot \partial_X \delta\acute{S}^{< {\text{NLO}}}(l,X) 
 -\cp l^\mu\partial^\nu_l F_{\mu\nu}\delta\acute{S}^{< {\text{LO}}}(l,X) \\
&= -\frac{\cp}{4}\biggl(\partial_{l\alpha}(\partial^\alpha_XF^{\mu\nu}) \sigma_{\mu\nu} S^{<\text{(eq)}}(l)\\
&~~~-iF^{\mu\nu}[\delta\acute{S}^{< {\text{LO}}}(l,X),\sigma_{\mu\nu}] \\
&~~~+\left\{\frac{\partial_l\cdot\partial_X}{2}F^{\mu\nu}\right\}[S^{<\text{(eq)}}(l),\sigma_{\mu\nu}]\biggr).
\end{split}
\end{align}
Since $\delta \acute{S}^{<{\text{LO}}}$ is linear in terms of $F_{\mu\nu}$, we see that $\delta \acute{S}^{<{\text{NLO}}}$ has quadratic component.
To obtain the linear and the quadratic components of $\delta \acute{S}^{<{\text{NLO}}}$ separately, we expand it as $\delta \acute{S}^{<{\text{NLO}}}=\delta \acute{S}^{<{\text{NLO}}}_1+\delta \acute{S}^{<{\text{NLO}}}_2$, where $\delta \acute{S}^{<{\text{NLO}}}_1$ ($\delta \acute{S}^{<{\text{NLO}}}_2$) is linear (quadratic) component.

$\delta \acute{S}^{<{\text{NLO}}}_1$ follows
\begin{align}
\label{eq:KB-NLO-1st}
\begin{split} 
&l\cdot \partial_X \delta\acute{S}^{< {\text{NLO}}}_1(l,X) \\
&= -\frac{\cp}{4}(\partial^\alpha_XF^{\mu\nu})\partial_{l\alpha}
\biggl( \sigma_{\mu\nu} S^{<\text{(eq)}}(l) 
+\frac{1}{2}[S^{<\text{(eq)}}(l),\sigma_{\mu\nu}]\biggr).
\end{split}
\end{align}
It is known that this equation can be rewritten in the form of the Vlasov equation with the term corresponding to the Berry phase~\cite{Chen:2012ca, Son:2012zy, Son:2012wh, Stephanov:2012ki}.
We can obtain the CME current from this equation, as is done in Appendix~\ref{app:CME}.
Here let us discuss the order of magnitude of $l$ that is relevant to our analysis.
As can be seen from Eq.~(\ref{eq:Pi-CME-1}), the current contains integral that has the form of $\int^\infty_0 d|\vl| [\exp\{\beta(|\vl|\mp\mu_{L/R})\}+1]^{-1}$, and the dominant contribution to the integral comes from the region $|\vl|\sim T$.
Since $\partial_s$ corresponds to $k-\cp A$ via the Wigner transformation, we confirm that $\partial_s\sim T$, which was assumed before, by using $\cp A\ll T$.

From Eq.~(\ref{eq:KB-NLO}), the quadratic component of $\delta \acute{S}^{<{\text{NLO}}}$ follows 
\begin{align}
\label{eq:KB-NLO-quadratic}
\begin{split} 
&l\cdot \partial_X \delta\acute{S}^{< {\text{NLO}}}_2(l,X) 
 -\cp l^\mu\partial^\nu_l F_{\mu\nu}\delta\acute{S}^{< {\text{LO}}}(l,X) \\
&= i\frac{\cp}{4}F^{\mu\nu}[\delta\acute{S}^{< {\text{LO}}}(l,X),\sigma_{\mu\nu}] .
\end{split}
\end{align}
If we write $\delta \acute{S}^{<{\text{NLO}}}_2$ as 
\begin{align}
\begin{split}
\delta \acute{S}^{<{\text{NLO}}}_2&= 2\pi \delta(l^2)
[\theta(l^0)\{P_L\delta n^2_{L+}+P_R\delta n^2_{R+}\}(\vl,X) \\
&~~~+\theta(-l^0)\{P_L\delta n^2_{L-}+P_R\delta n^2_{R-}\}(-\vl,X)]\Slash{l},
\end{split}
\end{align}
 we see that this equation coincides with the Vlasov equation at the quadratic order, Eq.~(\ref{eq:Vlasov-quadratic}), by using Eq.~(\ref{eq:S-LO-distribution}). 

\begin{figure}[t]
\begin{center} 
\includegraphics[width=0.58\textwidth]{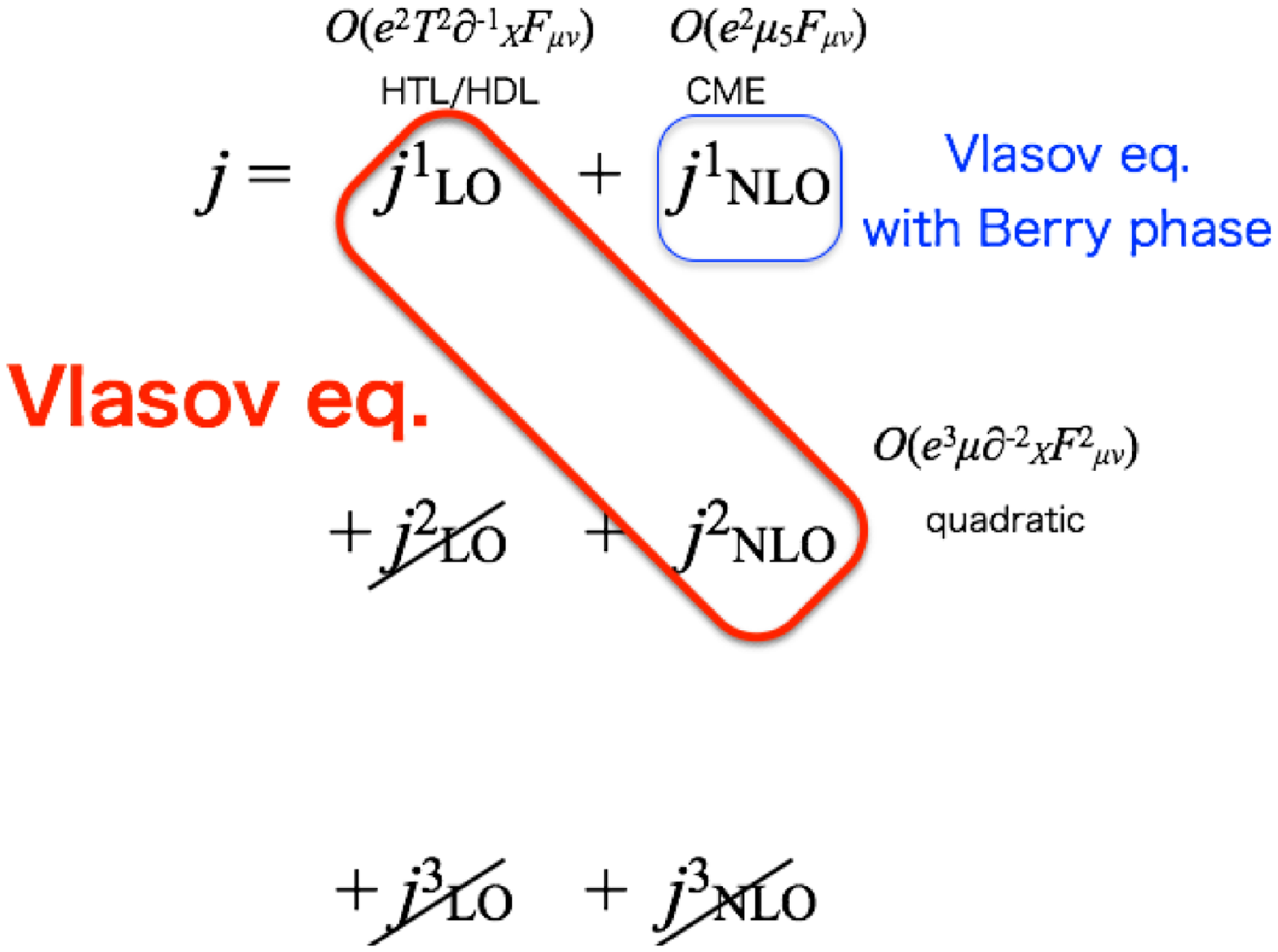}
\caption{Summary of the result of the analysis with the Kadanoff-Baym equation.}
\label{fig:summary}
\end{center}
\end{figure}

From Eq.~(\ref{eq:KB-NLO}), we also see that $\acute{S}$ does not contain more than two $F_{\mu\nu}$ at NLO, which implies that there are no induced currents that are higher than quadratic one.
Also, we see that the CME current and the quadratic current have the same order of magnitude, when $\cp F_{\mu\nu}\sim \partial^2_X$ is satisfied.
We note that the condition above is satisfied when we assume the conditions $\partial_X\sim\cp T$ and $F_{\mu\nu}\sim \cp T^2$, which is assumed in the derivation of the results of the HTL approximation from the Kadanoff-Baym equation~\cite{Blaizot:1992gn, Blaizot:2001nr}.
We briefly summarize the results in this subsection in Fig.~\ref{fig:summary}.

\section{quadratic current in HIC}
\label{ssc:HIC-vlasov}

In this section, we evaluate explicitly the quadratic current induced by a possible configuration of electromagnetic field realized in HIC.
It gives a demonstration of explicit calculation using Eq.~(\ref{eq:Vlasov-eqsult-quadratic}).
We will see how the quadratic current behaves differently in the local and nonlocal forms, and that the quadratic current can be comparable with the CME current.
We note that the latter property was also valid in the analysis done in Sec.~\ref{sec:KB}, although the parameters used in the present section does not satisfy the assumptions in Sec.~\ref{sec:KB}, as will be shown later.

\begin{figure}[t]
\begin{center} 
\includegraphics[width=0.3\textwidth]{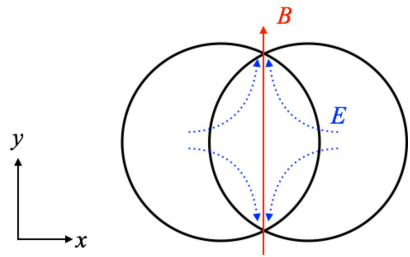}
\caption{Schematic picture of the possible electromagnetic field generated in HIC.
The dotted curve (blue) represents the electric field while the solid one (red) the magnetic field.}
\label{fig:HIC}
\end{center}
\end{figure}

As a configuration of electromagnetic field realized in HIC, we adopt the following one, which is similar to that used in Ref.~\cite{Hongo:2013cqa}:
\begin{align}
\label{eq:config-HIC-E}
\vE(X)&= \hat{y} E_0\frac{Y}{a} e^{-\vX^2/(2\sigma^2)}\theta(X_0),\\ 
\label{eq:config-HIC-B}
\vB(X)&= \hat{y} B_0 e^{-\vX^2/(2\sigma^2)}\theta(X_0),
\end{align}
where $X=(X_0,X,Y,Z)$.
Here, the transverse plane contains $x$ and $y$ axes, the magnetic field is parallel to $y$ axis, and the collision axis agrees with $z$ axis (see Fig.~\ref{fig:HIC}).
We note that the damping factor $e^{-X_0/b}$ in the electromagnetic field, which was present in Ref.~\cite{Hongo:2013cqa}, was approximated as $1$ here for simplicity.
This approximation is justified when the time we focus on is early enough.
For the parameters, we use the following values, which are used in Ref.~\cite{Hongo:2013cqa}:
\begin{align}
\label{eq:HIC-parameter}
\begin{split}
\cp E_0&= 2.0\times 10^{-2}~({\text{GeV}})^2,\\
\cp B_0&= 8.0\times 10^{-2}~({\text{GeV}})^2,\\
\sigma&= 4.0~{\text{fm}},\\
a&= 1.0~{\text{fm}},\\
\mu&=\mu_5=10~{\text{MeV}},\\
e&= 0.3.
\end{split}
\end{align}

Before doing the explicit evaluation, let us compare the order of magnitude of $\partial_X$ with that of $\tau^{-1}$.
The spatial dependence of the electromagnetic field is determined by the parameter $\sigma$, so $\partial_X\sim \sigma^{-1}=50$ MeV.
To evaluate $\tau$, we use the result of lattice calculation of electrical conductivity~\cite{Amato:2013naa, Gupta:2003zh}:
The electrical conductivity in the calculation where the up, down, and strange quarks are taken into account reads $C^{-1}\sigma_e/T\simeq 0.3$ around $T=300$ MeV with $C\equiv \sum_f q^2_f$, where $q_f$ is the electromagnetic charge of the quark with flavor index $f$~\cite{Amato:2013naa}. 
In our computation, $C=e^2$, thus $\sigma_e\simeq 0.3 e^2T$.
By using Eq.~(\ref{eq:conductivity-tau}), we get 
\begin{align}
\label{eq:tau-estimate}
\tau^{-1}= \frac{m^2_D}{3\sigma_e}\simeq 111~{\text{MeV}},
\end{align}
 at $T=300$ MeV, by using Eq.~(\ref{eq:HIC-parameter}) and assuming $T\gg \mu$.
This result suggests that $\partial_X$ and $\tau^{-1}$ are comparable, and thus the both cases that the collision effect is important/negligible should be considered.
Thus, we use the expressions of the quadratic current in the both cases, namely Eqs.~(\ref{eq:Boltzmann-result-2nd-0th}), (\ref{eq:Boltzmann-result-2nd-1st}) and (\ref{eq:Vlasov-eqsult-quadratic}).

\begin{figure}[t]
\begin{center} 
\includegraphics[width=0.5\textwidth]{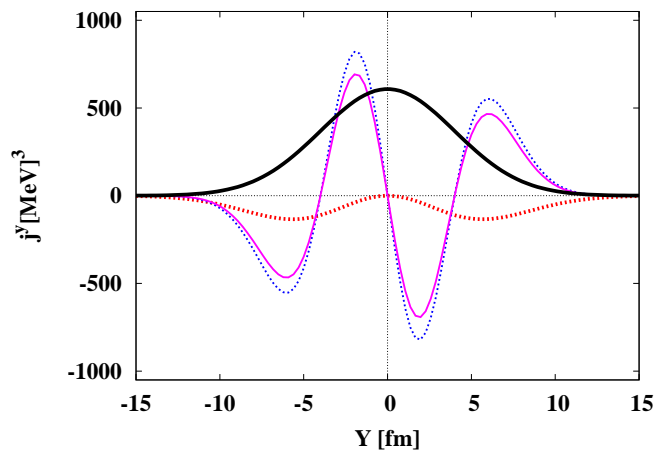}
\caption{The local and nonlocal quadratic currents, Eqs.~(\ref{eq:HIC-2nd-local-result}) and (\ref{eq:HIC-2nd-nonlocal-result}), and the local and nonlocal CME current, Eqs.~(\ref{eq:HIC-CME-local-result}) and (\ref{eq:HIC-CME-nonlocal-result}), as a function of $Y$.
The local currents are plotted after multiplying by $0.05$. 
We set $T=e\sigma$ and used the parameters Eq.~(\ref{eq:HIC-parameter}).
$0.05\times$Eq.~(\ref{eq:HIC-2nd-local-result}) is plotted with solid line (magenta), Eq.~(\ref{eq:HIC-2nd-nonlocal-result}) dotted line (blue), $0.05\times$Eq.~(\ref{eq:HIC-CME-local-result}) thick solid line (black), and Eq.~(\ref{eq:HIC-CME-nonlocal-result}) thick dotted line (red), respectively.
} 
\label{fig:HIC-current}
\end{center}
\end{figure}

Let us compare the quadratic currents in the both cases, to see how the nonlocal effect modifies the local current.
First, we evaluate the local current, Eqs.~(\ref{eq:Boltzmann-result-2nd-0th}), (\ref{eq:Boltzmann-result-2nd-1st}).
Since $\vE$ and $\vB$ are parallel, the hall current, Eq.~(\ref{eq:Boltzmann-result-2nd-0th}), is zero.
From Eqs.~(\ref{eq:Boltzmann-result-2nd-1st}) and (\ref{eq:config-HIC-E}), the $y$ component of the local current at $X_0>0$ reads
\begin{align}
\label{eq:HIC-2nd-local-result}
\begin{split} 
j^y_{2}(Y)&= \frac{2e^3\mu}{3\pi^2}
\left(\frac{E_0 }{a}\right)^2 \tau^3Y e^{-Y^2/\sigma^2}\left(\frac{Y^2}{\sigma^2}-1\right)
,
\end{split}
\end{align}
where we have focused on the region $X=Z=0$, in which the electromagnetic field is the stronger than other points.
Next, we evaluate the nonlocal current.
The $y$ component of the quadratic current at $\vX=(0,Y,0)$ is
\begin{align}
\begin{split}
j^y_2(Y)&= \frac{e^3\mu}{\pi^2} \int\frac{d\varOmega}{4\pi}v_y
\int^{X_0}_0 dt_1\int^{X_0-t_1}_0 dt_2 E^y(\alpha) \\
&~~~\times\bigl[E^y(\beta)(3v^2_y-1) \\
&~~~+(t_1+t_2)v_y(\nabla^y_XE^y(\beta)-v_y v^k \nabla^k_XE^y(\beta))\bigr] , 
\end{split}
\end{align}
from Eqs.~(\ref{eq:Vlasov-eqsult-quadratic}), (\ref{eq:config-HIC-E}), and (\ref{eq:config-HIC-B}).
To proceed the calculation analytically, from now on we focus on the case that $X_0$ is so small that $(X_0)^2\ll YX_0\ll\sigma^2$ is satisfied. 
From this condition, we have
\begin{align}
\label{eq:HIC-2nd-nonlocal-result}
\begin{split}
j^y_2(Y)&\simeq \frac{e^3\mu}{\pi^2} \frac{2}{15}\left(\frac{E_0}{a}\right)^2 (X_0)^3Y 
e^{-Y^2/{\sigma^2}} \left(\frac{Y^2}{\sigma^2}-1\right).
\end{split}
\end{align}
In Fig.~\ref{fig:HIC-current}, we plot Eqs.~(\ref{eq:HIC-2nd-local-result}) and (\ref{eq:HIC-2nd-nonlocal-result}).
In the plots we used Eqs.~(\ref{eq:HIC-parameter}) and (\ref{eq:tau-estimate}), and set $X_0=e\sigma$. 
We see that they have the same forms as functions of $Y$, but their orders of magnitude are very different:
The local current is larger than the nonlocal current by approximately a factor of 20.

Let us evaluate the CME current, to compare it with the quadratic current.
Due to the same reason as that for the quadratic current, we evaluate the CME current by using the local and nonlocal expression.
From Eq.~(\ref{eq:result-CME}), the nonlocal CME current reads
\begin{align}
\begin{split}
j^y_{\text{CME}}(X)&= -\frac{e^2}{2\pi^2}\mu_5 B_0 \biggl(\int\frac{d\varOmega}{4\pi}
e^{-(\vX^2-2\vv\cdot\vX X_0+(X_0)^2)/(2\sigma^2)} \\
&~~~-e^{-\vX^2/(2\sigma^2)}\biggr).
\end{split}
\end{align} 
If we focus on the region $X=Z=0$, we have
\begin{align}
\label{eq:HIC-CME-nonlocal-result}
\begin{split}
j^y_{\text{CME}}(Y)&= -\frac{e^2}{2\pi^2}\mu_5 B_0 e^{-Y^2/(2\sigma^2)}\\
&~~~\times\biggl[e^{-(X_0)^2/(2\sigma^2)}\frac{\sigma^2}{YX_0} 
\sinh\left(\frac{YX_0}{\sigma^2}\right) -1\biggr] \\
&\simeq -\frac{e^2}{2\pi^2}\mu_5 B_0 e^{-Y^2/(2\sigma^2)}\frac{1}{6}\left(\frac{YX_0}{\sigma^2}\right)^2, 
\end{split} 
\end{align}
where we have used $(X_0)^2\ll YX_0\ll \sigma^2$ in the last line.
We also evaluate the local CME current.
This current reads $\vj_{\text{CME}}(X)= e^2\mu_5\vB(X)/(2\pi^2)$~\cite{Fukushima:2008xe, Satow:2014lva}, which yields
\begin{align}
\label{eq:HIC-CME-local-result}
j^y_{\text{CME}}(Y)&=\frac{e^2}{2\pi^2}\mu_5 B_0 e^{-Y^2/(2\sigma^2)}.
\end{align}
Here we plot Eqs.~(\ref{eq:HIC-CME-nonlocal-result}) and (\ref{eq:HIC-CME-local-result}) in Fig.~\ref{fig:HIC-current}.
We see that they are comparable with the quadratic currents in both of the local and nonlocal expression.
This result suggests that, to analyze CME in HIC, it can be necessary to consider the quadratic current to subtract it from the total current, i.e., the quadratic current can be a background for the CME current.
We also see that, after averaging over $Y$,  the quadratic current vanishes while the CME current remains to be finite.
Thus, it is suggested that, to see the experimental effect of the quadratic current, we should see an observable quantity that is sensitive to fluctuation of the current $j^y(Y)$, not the one that is sensitive to the averaged current over $Y$.

Finally, we remark that the configuration of the electromagnetic field used in the present analysis does not satisfy the conditions assumed in Sec.~\ref{sec:KB}.
For example, one of the condition in Eq.~(\ref{eq:KB-condition}) is that $eF_{\mu\nu}/(T\partial_X)$ is much smaller than one, but this quantity is estimated as $\simeq 1.3$, which is comparable with one, by using Eq.~(\ref{eq:HIC-parameter}) around $T=300$ MeV.
Therefore, the result obtained in Sec.~\ref{sec:KB}, i.e., the nonlocal CME and quadratic currents have the same orders of magnitude, and the higher order currents in terms of $F_{\mu\nu}$ such as cubic one is much smaller than the quadratic current, can not be expected to be valid.
Nevertheless, our numerical result in this section shows that the former result is valid, so we could also expect the validity of the latter result.


\section{Summary and Concluding Remarks}
\label{sec:summary}

With HIC in mind, we analyzed the linear and the quadratic electromagnetic currents in terms of external electromagnetic field, in the two regimes:
In one regime, the scale of the inhomogeneity of electromagnetic field is so small that the collision effect is essentially important, and in the other regime, the inhomogeneity is so large that the collision effect is negligible.
In the former case, we listed all possible components of the linear and the quadratic currents in terms of the external electromagnetic field, and made an order estimate of each component by using the Boltzmann equation in the relaxation time approximation.
As a result, we found the magnitude of the strength of electromagnetic field with which the linear and the quadratic currents have the same order of magnitude.
In the latter case, we explicitly calculated the quadratic current by using the Vlasov equation, and found that the CME current and the quadratic current can have the same order of magnitude when Eq.~(\ref{eq:KB-condition}) is satisfied, by showing that, the Kadanoff-Baym equation at the NLO in the gradient expansion reproduces both the CME and the quadratic currents. 
Furthermore, we showed that there are no currents that are higher than the quadratic, e.g., cubic or quartic, do not appear in the analysis at the NLO. 
We emphasize that, as far as we know, these analyses are the first systematic studies on nonlinear electromagnetic response in the quark-gluon plasma. 
We also demonstrated that the quadratic current can have the same order of magnitude as that of CME current, by using a possible field configuration realized in HIC.

The results in this paper suggest that the quadratic current is the most sensitive term to $\mu$, so it could be useful to analyze the experimental effect of this current in HIC, in order to measure indirectly $\mu$ realized in HIC.
Especially, the low-energy scan done in Relativistic Heavy Ion Collider can be relevant since $\mu$ is expected to be relatively large. 
Also, the results suggest that, when the configuration of the electromagnetic field satisfies $\partial_X\sim\cp T$ and $F_{\mu\nu}\sim \cp T^2$, we can neglect the currents that are higher order than the quadratic one.

In this work, we computed the current around the thermal equilibrium state.
However, in HIC, the system expands, so taking into account this effect is one way to proceed the analysis further.
If we consider this effect, the distribution function becomes anisotropic, and thus the terms in Eqs.~(\ref{eq:list-1st}) and (\ref{eq:list-2nd}) that did not appear in Eqs.~(\ref{eq:Boltzmann-result-1st}) and (\ref{eq:Boltzmann-result-2nd-1st}), are expected to appear.
Also, the flow vector appears as a vector quantity with which we can construct the current, so we expect that there appears more terms~\cite{Gursoy:2014aka} in the current than those in our paper. 

Another way of improving the analysis in this paper is, to calculate the next-to-next-to-leading order terms with the Kadanoff-Baym equation.
In such analysis, we expect that the gradient expansion is more difficult to apply, and the HTL resummation~\cite{Pisarski:1988vd} becomes necessary due to the following reason:
As was discussed before, the dominant contribution to the current at the NLO comes from the region $|\vl|\sim T$.
By contrast, in the next-to-next-to-leading order calculation, we expect that integral like $\int^\infty_0 d|\vl| |\vl|^{-1} [\exp\{\beta(|\vl|\mp\mu_{L/R})\}+1]^{-1}$
appears instead of $\int^\infty_0 d|\vl| [\exp\{\beta(|\vl|\mp\mu_{L/R})\}+1]^{-1}$, and this integral contains infrared singularity.
After removing the singularity with the HTL resummation, which generates the infrared cutoff that is of order Debye mass, the dominant contribution comes from the region $|\vl|\sim \cp T$.
In this case, the assumption $\partial_X\ll T$, which justifies the gradient expansion, is replaced with $\partial_X\ll \cp T$, so the gradient expansion is more difficult to apply.

Finally, we remark that the calculation of the quadratic current in this paper is also relevant to analysis of photon splitting process~\cite{Adler:1970gg} induced by finite density, since the quadratic current contains the information of the three-point function of the photon~\cite{Blaizot:2001nr, Blaizot:1992gn}.
We leave the investigation of the photon splitting process in future work~\cite{future-photon-splitting}. 

\section*{Acknowledgements}
The author thanks M. Hongo, D.~Kharzeev, J.~Liao, L.~Mclerran, and A. Monnai for inspiring and fruitful discussion. 
He is indebted to N.~Yamamoto for collaboration in the beginning of this project and valuable comments. 
He is supported by JSPS Strategic Young Researcher Overseas Visits Program for Accelerating Brain Circulation (No. R2411). 

\appendix
\section{CME current calculated with Kadanoff-Baym equation}
\label{app:CME}
In this appendix, we derive Eq.~(\ref{eq:CME-Pi}) from Eq.~(\ref{eq:KB-NLO-1st}).
By using Eq.~(\ref{eq:propagator-eq}), Eq.~(\ref{eq:KB-NLO-1st}) becomes
\begin{align}
\label{eq:app-1}
\begin{split} 
l\cdot \partial_X \delta\acute{S}^{< {\text{NLO}}}_1(l,X) 
&= -i\frac{\cp}{8} (\partial_{l}\cdot\partial_XF^{\mu\nu})
\gamma_\nu\Slash{l}\gamma_\mu \rho^{0}(l) \\
&~~~\times(n^L+n^R+\gamma^5(n^L-n^R))(l^0). 
\end{split}
\end{align}
We note that this expression does not vanish if we multiply by $\Slash{l}$ from the right, in contrast to $\delta \acute{S}^{<{\text{LO}}}$ and $\delta \acute{S}^{<{\text{NLO}}}_2$. 
It reflects the fact that Eq.~(\ref{eq:app-1}) can not be written in the form of the Vlasov equation without the Berry phase term.
From this equation, we have
\begin{align}
\begin{split}
&l\cdot\partial_X \Tr[\gamma_\alpha \delta\acute{S}^{< {\text{NLO}}}_1(l,X)] \\
&= -\frac{\cp}{2} (\partial_{l}\cdot\partial_XF^{\mu\nu}) \epsilon_{\alpha \nu\beta\mu} l^\beta
\rho^0(l) (n^L-n^R)(l^0) .
\end{split}
\end{align}
The vector current is given by Eq.~(\ref{eq:current-vector-S}), so 
\begin{align} 
\label{eq:app-2}
\begin{split} 
&j^i_{\text{CME}}(p)\\
&~~~= \frac{\cp^2}{2} \int\frac{d^4l}{(2\pi)^4}
\frac{\epsilon_{i\nu\beta\mu} p^2}{(l\cdot p)^2}F^{\mu\nu}(p) l^\beta \rho^0(l) (n^L-n^R)(l^0)
\end{split}
\end{align}
in momentum space.
Here we did partial integration.
By using 
\begin{align}
j^i_{\text{CME}}(p)&= \varPi^{i\nu}_{R}(p) A_\nu(p),
\end{align}
 which is Fourier-transformed Eq.~(\ref{eq:CME-formal}), we get
\begin{align}
\label{eq:Pi-CME-1}
\begin{split}
\varPi_{ij}(p)
&= \frac{i\cp^2}{2\pi^2} \epsilon^{ijk} \int^\infty_0 d|\vl|\int\frac{d\varOmega}{4\pi}
\sum_{s=\pm 1} \frac{s|\vl|}{2}\frac{p^2}{(l\cdot p)^2}\\
&~~~\times (p^0l^k-p^kl^0)(n^L-n^R)(l^0)
\end{split} 
\end{align}
where $l^0=s|\vl|$.
This expression agrees with Eq.~(\ref{eq:CME-Pi}) after performing the integrations.

\section{CME current in coordinate space}
\label{app:CME-coordinate}
In this appendix, we derive Eq.~(\ref{eq:result-CME}).
By using Eq.~(\ref{eq:CME-Pi}), the current in the momentum space is given by
\begin{align}
\begin{split}
\vj_{\text{CME}}(p)&= \frac{e^2}{2\pi^2}\mu_5
\left(1-\frac{p^2_0}{|\vp|^2}\right) \\
&~~~\times\left(1+\frac{p_0}{2|\vp|}\ln \frac{p_0-|\vp|}{p_0+|\vp|}\right) \vB(p).
\end{split}
\end{align}
By using the Bianchi identity, $p^0\vB(p)=\vp\times\vE(p)$, we arrive at
\begin{align}
\label{eq:CME-p-2}
\begin{split}
\vj_{\text{CME}}(p)&= \frac{e^2\mu_5}{2\pi^2} \biggl[\vB(p)+\int\frac{d\varOmega}{4\pi}\frac{p^0}{p\cdot v} 
\left(\vv\times\vE(p)-\vB(p)\right)\biggr] .
\end{split}
\end{align}
We can switch to the coordinate space, by doing the Fourier transformation in Eq.~(\ref{eq:CME-p-2}).
Equation~(\ref{eq:CME-p-2}) can be rewritten as Eq.~(\ref{eq:result-CME}), by using 
\begin{align}
\frac{1}{p\cdot v}&= -i\int^\infty_0 dt e^{ip\cdot v t} .
\end{align}



\begin{thebibliography}{99}
  
\bibitem{elemag-HIC}  
  V.~Skokov, A.~Y.~.Illarionov and V.~Toneev,
  Int.\ J.\ Mod.\ Phys.\ A {\bf 24}, 5925 (2009)
  [arXiv:0907.1396 [nucl-th]];
  V.~Voronyuk, V.~D.~Toneev, W.~Cassing, E.~L.~Bratkovskaya, V.~P.~Konchakovski and S.~A.~Voloshin,
  Phys.\ Rev.\ C {\bf 83}, 054911 (2011)
  [arXiv:1103.4239 [nucl-th]];
  W.~-T.~Deng and X.~-G.~Huang,
  Phys.\ Rev.\ C {\bf 85}, 044907 (2012)
  [arXiv:1201.5108 [nucl-th]];
  L.~McLerran and V.~Skokov,
  arXiv:1305.0774 [hep-ph].

\bibitem{KW}
D.~E.~Kharzeev and H.~J.~Warringa,
  Phys.\ Rev.\ D {\bf 80}, 034028 (2009)
  [arXiv:0907.5007 [hep-ph]].

\bibitem{Fukushima:2008xe} 
  K.~Fukushima, D.~E.~Kharzeev and H.~J.~Warringa,
  Phys.\ Rev.\ D {\bf 78}, 074033 (2008)
  [arXiv:0808.3382 [hep-ph]].
For review articles, see 
  D.~E.~Kharzeev,
  Annals Phys.\  {\bf 325}, 205 (2010)
  [arXiv:0911.3715 [hep-ph]];
  K.~Fukushima,
  Lect.\ Notes Phys.\  {\bf 871}, 241 (2013)
  [arXiv:1209.5064 [hep-ph]];
  Prog.\ Theor.\ Phys.\ Suppl.\  {\bf 193}, 15 (2012);
  D.~Kharzeev, K.~Landsteiner, A.~Schmitt and H.~-U.~Yee,
  Lect.\ Notes Phys.\  {\bf 871}, 1 (2013).

\bibitem{Satow:2014lva} 
  D.~Satow and H.~-U.~Yee,
  arXiv:1406.1150 [hep-ph].



\bibitem{elemag-effectOnHIC} 
  M.~Asakawa, A.~Majumder and B.~Muller,
  Phys.\ Rev.\ C {\bf 81}, 064912 (2010)
  [arXiv:1003.2436 [hep-ph]].
  K.~Tuchin,
  Phys.\ Rev.\ C {\bf 82}, 034904 (2010)
  [Erratum-ibid.\ C {\bf 83}, 039903 (2011)]
  [arXiv:1006.3051 [nucl-th]].
  K.~Tuchin,
  Adv.\ High Energy Phys.\  {\bf 2013}, 490495 (2013)
  [arXiv:1301.0099].
  D.~E.~Kharzeev, L.~D.~McLerran and H.~J.~Warringa,
  Nucl.\ Phys.\ A {\bf 803}, 227 (2008)
  [arXiv:0711.0950 [hep-ph]].
  J.~Liao, V.~Koch and A.~Bzdak,
  Phys.\ Rev.\ C {\bf 82}, 054902 (2010)
  [arXiv:1005.5380 [nucl-th]].
  
\bibitem{Gursoy:2014aka} 
  U.~Gursoy, D.~Kharzeev and K.~Rajagopal,
  Phys.\ Rev.\ C {\bf 89}, 054905 (2014)
  [arXiv:1401.3805 [hep-ph]].
 
\bibitem{Hirono:2012rt} 
  Y.~Hirono, M.~Hongo and T.~Hirano,
  arXiv:1211.1114 [nucl-th].

\bibitem{Hongo:2013cqa} 
  M.~Hongo, Y.~Hirono and T.~Hirano,
  arXiv:1309.2823 [nucl-th].
  
\bibitem{Abelev:2009ac} 
  B.~I.~Abelev {\it et al.}  [STAR Collaboration],
  Phys.\ Rev.\ Lett.\  {\bf 103}, 251601 (2009)
  [arXiv:0909.1739 [nucl-ex]];
  B.~I.~Abelev {\it et al.}  [STAR Collaboration],
  Phys.\ Rev.\ C {\bf 81}, 054908 (2010)
  [arXiv:0909.1717 [nucl-ex]];
  B.~Abelev {\it et al.}  [ALICE Collaboration],
  Phys.\ Rev.\ Lett.\  {\bf 110}, 012301 (2013)
  [arXiv:1207.0900 [nucl-ex]];
  G.~Wang [STAR Collaboration],
  Nucl.\ Phys.\ A {\bf 904-905}, 248c (2013)
  [arXiv:1210.5498 [nucl-ex]];
  L.~Adamczyk {\it et al.}  [STAR Collaboration],
  arXiv:1302.3802 [nucl-ex];
  L.~Adamczyk {\it et al.}  [STAR Collaboration],
  Phys.\ Rev.\ C {\bf 89}, 044908 (2014)
  [arXiv:1303.0901 [nucl-ex]].  
  
   
\bibitem{Buividovich:2010tn} 
  P.~V.~Buividovich, M.~N.~Chernodub, D.~E.~Kharzeev, T.~Kalaydzhyan, E.~V.~Luschevskaya and M.~I.~Polikarpov,
  Phys.\ Rev.\ Lett.\  {\bf 105}, 132001 (2010)
  [arXiv:1003.2180 [hep-lat]].  

  \bibitem{Blaizot:1992gn} 
  J.~-P.~Blaizot and E.~Iancu,
  Nucl.\ Phys.\ B {\bf 390}, 589 (1993);
  Phys.\ Rev.\ Lett.\  {\bf 70}, 3376 (1993)
  [hep-ph/9301236];
  Nucl.\ Phys.\ B {\bf 417}, 608 (1994)
  [hep-ph/9306294].
  
\bibitem{Blaizot:2001nr} 
  J.~-P.~Blaizot and E.~Iancu,
  Phys.\ Rept.\  {\bf 359}, 355 (2002)
  [hep-ph/0101103].
   
\bibitem{Frenkel:1989br} 
  J.~Frenkel and J.~C.~Taylor,
  Nucl.\ Phys.\ B {\bf 334}, 199 (1990);
  E.~Braaten and R.~D.~Pisarski,
  Nucl.\ Phys.\ B {\bf 339}, 310 (1990).

\bibitem{Altherr:1992mf} 
  T.~Altherr and U.~Kraemmer,
  Astropart.\ Phys.\  {\bf 1}, 133 (1992);
  H.~Vija and M.~H.~Thoma,
  Phys.\ Lett.\ B {\bf 342}, 212 (1995)
  [hep-ph/9409246];
  C.~Manuel,
  Phys.\ Rev.\ D {\bf 53}, 5866 (1996)
  [hep-ph/9512365].

\bibitem{Son:2012zy} 
  D.~T.~Son and N.~Yamamoto,
  Phys.\ Rev.\ D {\bf 87}, 085016 (2013)
  [arXiv:1210.8158 [hep-th]].

\bibitem{Chen:2012ca} 
  J.~-W.~Chen, S.~Pu, Q.~Wang and X.~-N.~Wang,
  Phys.\ Rev.\ Lett.\  {\bf 110}, 262301 (2013)
  [arXiv:1210.8312 [hep-th]].
  
\bibitem{Son:2012wh} 
  D.~T.~Son and N.~Yamamoto,
  Phys.\ Rev.\ Lett.\  {\bf 109}, 181602 (2012)
  [arXiv:1203.2697 [cond-mat.mes-hall]].

\bibitem{Akamatsu:2013pjd} 
  Y.~Akamatsu and N.~Yamamoto,
  Phys.\  Rev.\  Lett.\  {\bf 111}, 052002 (2013)
  [arXiv:1302.2125 [nucl-th]];
  arXiv:1402.4174 [hep-th].

\bibitem{Arnold:2000dr} 
  P.~B.~Arnold, G.~D.~Moore and L.~G.~Yaffe,
  JHEP {\bf 0011}, 001 (2000)
  [hep-ph/0010177];
  JHEP {\bf 0305}, 051 (2003)
  [hep-ph/0302165].
  
\bibitem{Hosoya:1983xm} 
  A.~Hosoya and K.~Kajantie,
  Nucl.\ Phys.\ B {\bf 250}, 666 (1985);
  G.~Baym, H.~Monien, C.~J.~Pethick and D.~G.~Ravenhall,
  Phys.\ Rev.\ Lett.\  {\bf 64}, 1867 (1990);
  G.~D.~Moore and J.~-M.~Robert,
  hep-ph/0607172.

\bibitem{Huang:2013iia} 
  X.~-G.~Huang and J.~Liao,
  Phys.\ Rev.\ Lett.\  {\bf 110}, no. 23, 232302 (2013)
  [arXiv:1303.7192 [nucl-th]].

  
\bibitem{Son:2004tq} 
  D.~T.~Son and A.~R.~Zhitnitsky,
  Phys.\ Rev.\ D {\bf 70}, 074018 (2004)
  [hep-ph/0405216];
  M.~A.~Metlitski and A.~R.~Zhitnitsky,
  Phys.\ Rev.\ D {\bf 72}, 045011 (2005)
  [hep-ph/0505072].

\bibitem{Blaizot:1999xk} 
  J.~-P.~Blaizot and E.~Iancu,
  Nucl.\ Phys.\ B {\bf 557}, 183 (1999)
  [hep-ph/9903389].
  
\bibitem{Stephanov:2012ki} 
  M.~A.~Stephanov and Y.~Yin,
  Phys.\ Rev.\ Lett.\  {\bf 109}, 162001 (2012)
  [arXiv:1207.0747 [hep-th]].

\bibitem{Gupta:2003zh} 
  S.~Gupta,
  Phys.\ Lett.\ B {\bf 597}, 57 (2004)
  [hep-lat/0301006];
  G.~Aarts, C.~Allton, J.~Foley, S.~Hands and S.~Kim,
  Phys.\ Rev.\ Lett.\  {\bf 99}, 022002 (2007)
  [hep-lat/0703008 [HEP-LAT]];
  H.~-T.~Ding, A.~Francis, O.~Kaczmarek, F.~Karsch, E.~Laermann and W.~Soeldner,
  Phys.\ Rev.\ D {\bf 83}, 034504 (2011)
  [arXiv:1012.4963 [hep-lat]];
  A.~Francis and O.~Kaczmarek,
  Prog.\ Part.\ Nucl.\ Phys.\  {\bf 67}, 212 (2012)
  [arXiv:1112.4802 [hep-lat]];
  Y.~Burnier and M.~Laine,
  Eur.\ Phys.\ J.\ C {\bf 72}, 1902 (2012)
  [arXiv:1201.1994 [hep-lat]];
  O.~Kaczmarek and M.~M?ller,
  PoS LATTICE {\bf 2013}, 175 (2013)
  [arXiv:1312.5609 [hep-lat]];
  B.~B.~Brandt, A.~Francis, H.~B.~Meyer and H.~Wittig,
  JHEP {\bf 1303}, 100 (2013)
  [arXiv:1212.4200 [hep-lat]].

  
\bibitem{Amato:2013naa} 
  A.~Amato, G.~Aarts, C.~Allton, P.~Giudice, S.~Hands and J.~-I.~Skullerud,
  Phys.\ Rev.\ Lett.\  {\bf 111}, 172001 (2013)
  [arXiv:1307.6763 [hep-lat]].
  
\bibitem{Pisarski:1988vd} 
  R.~D.~Pisarski,
  Phys.\ Rev.\ Lett.\  {\bf 63}, 1129 (1989);
  E.~Braaten and R.~D.~Pisarski,
  Phys.\ Rev.\ Lett.\  {\bf 64}, 1338 (1990);
  Nucl.\ Phys.\ B {\bf 337}, 569 (1990);
  Phys.\ Rev.\ D {\bf 42}, 2156 (1990).
  
\bibitem{Adler:1970gg} 
  S.~L.~Adler, J.~N.~Bahcall, C.~G.~Callan and M.~N.~Rosenbluth,
  Phys.\ Rev.\ Lett.\  {\bf 25}, 1061 (1970);
  S.~L.~Adler,
  Annals Phys.\  {\bf 67}, 599 (1971);
  V.~N.~Baier, A.~I.~Milshtein and R.~Z.~.Shaisultanov,
  Phys.\ Rev.\ Lett.\  {\bf 77}, 1691 (1996)
  [hep-th/9604028];
  S.~L.~Adler and C.~Schubert,
  Phys.\ Rev.\ Lett.\  {\bf 77}, 1695 (1996)
  [hep-th/9605035];
  J.~I.~Weise,
  Phys.\ Rev.\ D {\bf 69}, 105017 (2004);
  G.~Brodin, M.~Marklund, B.~Eliasson and P.~K.~Shukla,
  Phys.\ Rev.\ Lett.\  {\bf 98}, 125001 (2007)
  [astro-ph/0702364].
  
\bibitem{future-photon-splitting}
K.~Hattori, D.~Satow, and N.~Yamamoto,
in preparation.

\end{thebibliography}
\end{document}